\documentstyle[12pt]{article}
\begin{document}
\hfill{NCKU-HEP-97-02}\par
\hfill{NTU-TH-97-07}
\vfill

\centerline{\large\bf Nonfactorizable soft gluons in}
\centerline{\large\bf nonleptonic heavy meson decays}
\vskip 1.0cm
\centerline{Hsiang-nan Li}\par
\centerline{Department of Physics, National Cheng-Kung University,}
\centerline{Tainan, Taiwan, R.O.C.}
\vskip 0.3cm
\centerline{B. Tseng}\par
\centerline{Department of Physics, National Taiwan University,}
\centerline{Taipei, Taiwan, R.O.C.}
\vskip 1.0cm
\centerline{\today }
\vskip 2.0 cm

\vskip 1.0 cm
\centerline{\bf Abstract}

We include nonfactorizable soft gluon corrections into the perturbative QCD 
formalism for exclusive nonleptonic heavy meson decays, which combines 
factorization theorems and effective field theory. These corrections are 
classified according to their color structures, and exponentiated separately
to complete the Sudakov resummation up to next-to-leading logarithms. 
The nonfactorizable contributions in nonleptonic decays are 
clearly identified in our formalism, and found to be positive for bottom 
decays and negative for charm decays. Our analysis confirms that the
large-$N_c$ approximaton is applicable to charm decays, but not to bottom
decays, consistent with the phenomenological implications of experimental
data. The comparision of our predictions with those from QCD sum rules is
also made.

\vfill

\newpage
\centerline{\bf I. INTRODUCTION}
\vskip 0.5cm

Nonleptonic heavy meson decays are more difficult to analyze compared to 
semileptonic decays, because they involve complicated strong interactions. 
A naive perturbative QCD (PQCD) formalism \cite{LY1}, considering dynamics 
between the large typical scale $t$ of decay processes, which is of order of
the heavy meson mass, and the hadronic scale of order $\Lambda_{\rm QCD}$, 
is appropriate only for semileptonic decays. For nonleptonic decays, a more 
sophisticated formalism must be developed, which further includes dynamics 
between the $W$ boson mass $M_W$ and the typical scale $t$. Recently, such 
a modified PQCD framework has been proposed \cite{CL}, in which nonleptonic 
decay rates are factorized into the convolution of a ``harder" function
characterized by $M_W$, a hard subamplitude by $t$, and meson wave functions 
by the hadronic scale. The renormalization-group (RG) method and the 
resummation technique \cite{CS} are then applied to organize large single 
logarithms $\ln(M_W/t)$ and double logarithms $\ln(t/\Lambda_{\rm QCD})$ 
contained in the perturbative expansions of these convolution functions into 
Wilson coefficients \cite{Buras} and Sudakov factors \cite{LY1,L1}, 
respectively. In the previous analysis of \cite{WYL} we have considered
nonfactorizable contributions to the hard subamplitude. However, 
nonfactorizable soft gluon effects, which cannot be absorbed into 
meson wave functions, were neglected because of their complicated color 
flows. In this paper we shall take into account these soft corrections,
which produce single logarithms, systematically in the modified PQCD 
formalism, and improve the accuracy of the resummation up to next-to-leading 
logarithms. 

The simplest and most widely adopted approach to exclusive nonleptonic 
heavy meson decays is the Bauer-Stech-Wirbel (BSW) model \cite{BSW} based 
on the factorization hypothesis, in which decay rates are expressed in 
terms of various hadronic transition form factors multiplied by some 
coefficients. The coefficients of the form factors corresponding to 
external $W$ boson emission and to internal $W$ boson emission
are $a_1=c_1+c_2/N_c$ and $a_2=c_2+c_1/N_c$, respectively, $N_c=3$ being the 
number of colors. Here $c_i$ are the Wilson coefficients appearing in the 
effective Hamiltonian,
\begin{equation}
H_{\rm eff} = \frac{4G_{F}}{\sqrt{2}} V_{cb} V^{\ast}_{ud} 
 [ \, c_{1}(\mu) O_{1} + c_{2}(\mu) O_{2} \, ]\;,
\label{eff}
\end{equation}
which satisfy the matching conditions $c_1(M_W)=1$ and $c_2(M_W)=0$. 
$O_{1}  =  (\bar{c}_{L}\gamma_{\mu} b_{L})
 (\bar{d}_{L}\gamma^{\mu} u_{L})$ and 
$O_{2}  =  (\bar{d}_{L}\gamma_{\mu} b_{L})            
(\bar{c}_{L}\gamma^{\mu} u_{L})$ are the local four-fermion operators
associated with bottom decays. The form factors may be related to each other
by heavy quark symmetry, and be modelled by different ansatz. The 
nonfactorizable contributions which cannot be expressed in terms of hadronic 
form factors, and the nonspectator contributions from $W$ 
boson exchanges are neglected. In this way the BSW method avoids 
complicated QCD dynamics. 

Though the BSW model is simple and gives predictions in fair agreement with 
experimental data, it encounters several difficulties. It has been observed
that the naive factorization is incompatible with experimental data 
for color-suppressed decay modes \cite{BSW}. 
Therefore, the large $N_c$ limit of $a_{1,2}$, {\it ie.} the 
choice $a_{1}=c_{1}(M_c)\approx 1.26$ and $a_{2}=c_{2}(M_c)\approx -0.52$,
with $M_c$ the $c$ quark mass, must be employed in order to explain the data 
of charm decays \cite{BSW}. 
However, the same limit of $a_{1}=c_{1}(M_b)\approx 1.12$ and 
$a_{2}=c_{2}(M_b)\approx -0.26$, $M_b$ being the $b$ quark mass, does not 
apply to the bottom case. Even after including the $c_{1,2}/N_c$ term such
that $a_1=1.03$ and $a_2=0.11$, the BSW predictions are still insufficient
to match the data. To overcome this difficulty, parameters $\chi$, denoting
the nonfactorizable contributions which are suppressed in the large $N_c$
limit \cite{largeN}, have been introduced \cite{HYC1}. They lead to the
effective coefficients
\begin{equation}
a^{\rm eff}_1=c_1+c_2\left(\frac{1}{N_c}+\chi_1\right)\;,\;\;\;\;
a^{\rm eff}_2=c_2+c_1\left(\frac{1}{N_c}+\chi_2\right)\;.
\end{equation}
The parameters $\chi$ should be 
negative for charm decays, canceling the color-suppressed 
term $1/N_c$, and be positive for bottom decays in order to account for
the observed constructive interference in $B\to D^{(*)}\pi$ decays. 
A phenomenological extraction \cite{HYC1} from the CLEO data
\cite{A} gave
\begin{eqnarray}
& &\chi_1(D\to {\bar K}\pi)\approx \chi_2(D\to {\bar K}\pi)
\approx -0.36\;,
\nonumber \\
& &\chi_1(B\to D\pi)\approx 0.05\;,\;\;\;\;
\chi_2(B\to D\pi)\approx 0.11\;.
\label{chii}
\end{eqnarray}
We shall demonstrate that our predictions for the nonfactorizable
contributions are consistent with the above values of $\chi$.

The rule of discarding the $1/N_c$ corrections \cite{BGR}, {\it i.e.}
the large $N_c$ approximation, found its
dynamical origin in the analyses based on QCD sum rules \cite{BS87}: 
$\chi$ for charm decays are indeed negative, and cancel the term $1/N_c$ 
roughly. However, the sum rule analyses also predicted negative $\chi$ for 
the bottom decay ${\bar B}^0\to D^+\pi^-$ in a certain kinematic limit
\cite{BS93}, and are thus in conflict with the phenomenological implications.
Hence, the mechanism responsible for the sign change has not been understood
completely, and remains a challenging subject in weak nonleptonic decays
\cite{BS93}. In our formalism with the soft gluons taken into account, the
nonfactorizable contributions can be clearly identified. We shall investigate
how the soft corrections modify the previous predictions for heavy meson
decays \cite{CL,WYL}, and try to explore the dynamical origin for the sign
change of the nonfactorizable contributions in bottom and charm decays in a
unified viewpoint.

In Sect. II we briefly review the derivation of the three-scale PQCD
factorization formulas for heavy meson decays, and include the
nonfactorizable soft gluon effects into the formulas. The numerical results,
along with a detailed comparision with those from the phenomenological and
QCD sum rule analyses, are presented in Sect. III. Section IV
is our conclusion.
\vskip 1.0cm

\centerline{\bf II. SOFT GLUON CORRECTIONS}
\vskip 0.5cm

The construction of the modified PQCD formalism is as follows. Before
including QCD corrections, nonleptonic heavy meson decays are described
by a full Hamiltonian of four-quark current-current operators, such as
\begin{equation}
H= \frac{4G_{F}}{\sqrt{2}} V_{cb} V^{\ast}_{ud} 
   (\bar{c}_{L}\gamma_{\mu} b_{L})
            (\bar{d}_{L}\gamma^{\mu} u_{L})
\label{FourFer}
\end{equation}
for the $B\to D^{(*)}\pi(\rho)$ decays. Consider one-loop corrections
to a tree-level heavy quark decay amplitude without integrating out the $W$
boson. The amplitude is ultraviolet finite because of the current
conservation (a conserved current is not renormalized) and the presence of
the $W$ boson line. However, these corrections give rise to infrared
divergences, when the radiative gluons are soft or collinear to the involved
light quarks. We classify the higher-order diagrams into the reducible and
irreducible types \cite{LS}: The former contains double logarithms from the
overlap of soft and collinear divergences, while the latter contains only
single soft logarithms.

We factorize the infrared sensitive contributions according to Fig.~1(a) 
first, which are collected by the appropriate eikonal approximation for 
quark propagators. The diagram in the second parentheses is absorbed into a 
meson wave function $\phi(b,\mu)$, if it is two-particle reducible, or into 
a soft function $U(b,\mu)$, if it is two-particle irreducible as exemplified 
by Fig.~1(a). Both $\phi$ and $U$ depend on a renormalization scale $\mu$, 
since the eikonal approximation brings in ultraviolet divergences. The 
variable $b$, conjugate to the transverse momentum $k_T$ carried by a 
valence quark, can be regarded as the spatial extent of the meson. $1/b$ is
then the hadronic scale, which will play the role of an infrared cutoff for 
loop integrals below. 

The diagrams in the first parentheses of Fig.~1(a) involve scales above 
$1/b$, and need further factorization. We express the full $O(\alpha_s)$ 
diagram into two terms, with the first term obtained by shrinking the $W$ 
boson line into a point, and the second term being the difference between 
the full diagram and the first term. Evidently, the former is characterized 
by the scale $t\ll M_W$ introduced before, and the latter by momenta of order
$M_W$. We factorize the contributions with the scale $M_W$ according to
Fig.~1(b), grouping it into a ``harder" function $H_r(M_W,\mu)$ (not an
amplitude). The remaining diagrams in the last parentheses, because of
operator mixing, correspond to the local four-fermion operators 
$O_{1}$ or $O_{2}$ in Eq.~(\ref{eff}). They are
absorbed into a hard decay subamplitude $H(t,\mu)$. Similarly, shrinking the 
$W$ boson line brings in ultraviolet divergences, and thus both $H_r$ and 
$H$ acquire a $\mu$ dependence.

Therefore, we arrive at the factorization formula for the decay amplitude
\begin{equation}
{\cal M}=H_r(M_W,\mu)\otimes H(t,\mu)\otimes \phi(b,\mu)\otimes U(b,\mu)\;,
\label{hpu}
\end{equation}
where $\otimes$ represents a convolution relation, since $t$ and $b$ will 
be integrated out at last. Assume that $\gamma_{H_r}$, $\gamma_\phi$, and 
$\gamma_U$ are the anomalous dimensions of $H_r$, $\phi$, and $U$, 
respectively. The anomalous dimension of $H$ is then 
$\gamma_H=-(\gamma_{H_r}+\gamma_\phi+\gamma_U)$, because a decay 
amplitude is ultraviolet finite as stated above. A RG treatment of 
Eq.~(\ref{hpu}) leads to
\begin{eqnarray}
{\cal M}&=&H_r(M_W,M_W)\otimes H(t,t)\otimes \phi(b,1/b)\otimes U(b,1/b)
\nonumber \\
& &\otimes
\exp\left[\int_{t}^{M_W}\frac{d{\bar\mu}}
{\bar\mu}\gamma_{H_r}(\alpha_s({\bar\mu}))\right]
\nonumber\\
& &\otimes
\exp\left[-\int_{1/b}^t\frac{d{\bar\mu}}
{\bar\mu}\left[\gamma_\phi(\alpha_s({\bar\mu}))
+\gamma_U(\alpha_s({\bar\mu}))\right]\right]\;,
\label{main}
\end{eqnarray}
which becomes explicitly $\mu$-independent. The two exponentials,
describing the two-stage evolutions from $M_W$ to $t$ and from $t$ to $1/b$, 
are the consequence of the summation of large $\ln(M_W/t)$ and 
$\ln(tb)$. The first exponential can be easily identified as the Wilson 
coefficient $c(t)$. Note that the Wilson coefficient appears as a 
convolution factor here, instead of a multiplicative factor in the 
effective Hamiltonian, Eq.~(\ref{eff}). In the conventional approach to 
exclusive nonleptonic heavy meson decays \cite{BSW}, which is based on 
Eq.~(\ref{eff}), a value of $\mu$ must be assigned, and thus an ambiguity is 
introduced \cite{LSW}. A merit of our formalism is then that it does not
suffer the scale setting ambiguity. 

Equation (\ref{main}) sums only the single logarithms. In fact, there exist 
the double logarithms $\ln^2(P^+b)$ in the meson wave function $\phi$, which 
arise from the overlap of collinear and soft divergences,
$P^+$ being the largest light-cone component of the meson momentum.
Hence, $\phi(b,\mu)$ in Eq.~(\ref{hpu}) should be replaced by 
\begin{equation}
\phi(P^+,b,\mu)=\phi(b,\mu)\exp[-s(P^+,b)]\;,
\label{sdc}
\end{equation}
where the exponential $e^{-s}$, the so-called Sudakov form factor, comes
from the resummation of the double logarithms. 
For the detailed derivation of Eq.~(\ref{sdc}) and the complete formula of 
$s$, refer to \cite{LY1,L1}. Below we shall approximate $H_r(M_W,M_W)$ by 
its lowest-order expression $H^{(0)}_r=1$, and evaluate $H(t,t)$ 
perturbatively, since all the large logarithms have been grouped into the 
exponents. For simplicity, we set the nonperturbative initial condition 
$U(b,1/b)$ of the RG evolution to unity due to the strong suppression at 
large $b$ \cite{BS}, and neglect the $b$ dependence in another 
nonperturbative initial condition $\phi(b,1/b)$. 

At the tree level, six types of diagrams contribute to the hard decay 
subamplitude of $B^-(P_1)\to D^0(P_2)\pi^-(P_3)$ as shown in Fig.~2, where
the momenta are chosen as 
\begin{equation}
P_1=\frac{M_B}{\sqrt{2}}(1,1,{\bf 0})\;,\;\;\;\;
P_2=\frac{M_B}{\sqrt{2}}(1,r^2,{\bf 0})\;,\;\;\;\;
P_3=\frac{M_B}{\sqrt{2}}(0,1-r^2,{\bf 0})\;,
\end{equation}
with $r=M_D/M_B$, $M_B$ $(M_D)$ being the $B$ $(D)$ meson mass. These
diagrams represent external $W$ emissions, if the four quark operators are 
$O_1$ in Figs.~2(a) and 2(b), and $O_2$ in Figs.~2(c)-2(f). They are 
internal $W$ emissions, if the four quark opeartors are $O_2$ in Figs.~2(a) 
and 2(b), and $O_1$ in Figs.~2(c)-2(f). At this level, only Figs.~2(e) and 
2(f) give nonfactorizable contributions. We shall argue that after including
the soft gluons, Figs.~2(c) and 2(d) become nonfactorizable. 

We discuss the color structure of soft gluon exchanges. If a soft gluon 
crosses the hard gluon vertex, the color structure is given by
\begin{eqnarray}
& &(T^a)_{b_1 b^{\prime}_1}(T^H)_{b^{\prime}_1 b_2}
(T^a)_{a_2 a^{\prime}_2}(T^H)_{a^{\prime}_2 a_1}
\nonumber\\
& &=\frac{1}{2}(T^H)_{a_2 b_2}(T^H)_{b_1 a_1}-
{1 \over 2N_c}(T^H)_{b_1 b_2}(T^H)_{a_2 a_1}\;,
\end{eqnarray}
where $T^a$ $(T^H)$ is the color matrix 
associated with the soft (hard) gluon. Contracted with 
$\delta_{a_1 b_1} \delta_{a_2 b_2}$ from the color-singlet initial- and 
final-state mesons, the first term, denoting an octet contribution, 
diminishes. The second term leads to a factor $-1/(2 N_c)$ without changing 
the original tree-level color flow. If a soft gluon does not cross the hard 
gluon vertex, it introduces a factor $T^aT^a={\cal C}_F=4/3$. 
It will be shown that each type of hard diagrams acquires different soft 
corrections, which must be organized separately.

For the external $W$ emissions, the soft function receives two contributions 
as shown in Fig.~3(a):
\begin{eqnarray}
U^{(e)}=-\frac{1}{2 N_c}[I(p_1,p_2) + I(k_1,k_2)]\;.
\label{ue}
\end{eqnarray}
Employing the eikonal approximation for the quark propagators, the loop 
integral $I$ is written as
\begin{eqnarray}
I(u,v)=-ig^{2} \mu^{ \epsilon} 
       \int \frac{d^{4-\epsilon} l}{(2 \pi)^{4-\epsilon}}
       \frac{u^{\alpha}}{u \cdot l}
       \frac{v^{\beta}}{v \cdot l}
       \frac{N_{\alpha \beta}}{l^2}\;, 
\end{eqnarray}
with the tensor
\begin{equation}
N_{\alpha \beta}= g_{\alpha \beta}
-\frac{n_\alpha l_\beta+n_\beta l_\alpha}{n\cdot l}
+n^2\frac{l_\alpha l_\beta}{(n\cdot l)^2}
\end{equation}
from the gluon propagator in the axial gauge $n\cdot A=0$, 
$n=(1,-1,{\bf 0})$ being a gauge vector. The other soft contributions
either vanish because of the color flow in Figs.~2(a) and 2(b), or cancel
by pairs in Figs.~2(c)-2(f). 

For convenience, the dimensionless vectors associated with the corresponding 
meson momenta are assigned as
\begin{eqnarray}
& &p_1=(1,1,{\bf 0})\;,\;\;\;\; p_2=(1,r^2,{\bf 0})\;,
\nonumber \\
& &k_1=(0,1,{\bf 0})\;,\;\;\;\; k_2=(1,0,{\bf 0})\;,
\end{eqnarray}
for the $b$ quark, the $c$ quark, the light valence quark in the $B$ meson,
and the light valence quark in the $D$ meson, respectively. The dimensionless 
vector associated with the pion is then $p_3=(0,1,{\bf 0})$. All the above 
kinematic variables can be copied to charm decays directly, such as 
$D\to {\bar K}\pi$, by choosing $r=M_K/M_D$, $M_K$ being the kaon mass. 
Varying $r$, we can study how the soft corrections modify the 
predictions for bottom and charm decays. 

Concentrating only on the pole terms, we obtain
\begin{eqnarray} 
I(p_1,p_2) & = & -\frac{\alpha_{s}}{\pi}\frac{1}{\epsilon}
[ f(p_1,p_2)- f(p_1,n)- f(p_2,n)+1]\;,
\label{uhh}\\
I(k_1,k_2) & = & -\frac{\alpha_{s}}{\pi}
          \frac{1}{\epsilon} \left[\ln\frac{k_1 \cdot k_2}{2}
-\frac{1}{2} \ln\frac{(k_1 \cdot n)^2}{n^2}
- \frac{1}{2}\ln\frac{ (k_2 \cdot n)^2}{n^2}+1 \right]\;, 
\label{ull}
\end{eqnarray}
with 
\begin{eqnarray}
f(u,v)&=&u\cdot v\int_{0}^{1} dx [x^2 u^2 +
 2 x(1-x){u \cdot v}+(1-x)^2 v^2]^{-1}\;.
\end{eqnarray}
A straightforward calculation gives
\begin{eqnarray} 
I(p_1,p_2) & = & -\frac{\alpha_{s}}{\pi}\frac{1}{\epsilon}
\left(\frac{4r^2}{r^4-1}\ln r+1\right)\;,
\\
I(k_1,k_2)&=& -\frac{\alpha_{s}}{\pi}\frac{1}{\epsilon}\;,
\end{eqnarray}
from which we extract the anomalous dimension
\begin{equation}
\gamma^{(e)}_U=-\frac{\alpha_{s}}{\pi}\frac{1}{N_c}
\left(\frac{2r^2}{r^4-1}\ln r+1\right)\;.
\end{equation}

For the internal $W$ emissions, the analysis is more complicated. Soft
corrections to Fig.~2(a) and 2(b) are similar to Eq.~(\ref{ue}) but
with $p_2$ and $k_2$ replaced by $p_3$, that is,
\begin{eqnarray}
U^{(i1)}=-\frac{1}{2 N_c}[I(p_1,p_3) + I(k_1,p_3)]\;.
\label{ui1}
\end{eqnarray}
Here $k_1$ should be chosen as $k_1=(1,0,{\bf 0})$ to render the 
second term meaningful. Other soft corrections, again, vanish because of the 
color flow. To Figs.~2(c) and 2(e), six soft gluon exchange diagrams 
contribute as shown in Fig.~3(b). The soft function is given by
\begin{eqnarray}
U^{(i2)}&=&-\frac{1}{2 N_c}[ I(p_1,p_2) - I(p_1,k_2) +
I(p_1,p_3) +I(k_1,p_3)]
\nonumber \\
& &+{\cal C}_F[I(k_2,p_3) -I(p_2,p_3)]\;.
\label{ui2}
\end{eqnarray}
Notice the minus signs in front of $I(p_1,k_2)$ and $I(p_2,p_3)$. A minus
sign appears, when the soft gluon attaches a quark and an antiquark 
\cite{BS}. If the $D$ meson is massless, pair cancellations occur between 
$I(p_1,p_2)$ and $I(p_1,k_2)$ in the first brackets, and between the two 
terms in the second brackets. $U^{(i2)}$ then reduces to $U^{(i1)}$. The 
soft function associated with Figs.~2(d) and 2(f) is 
\begin{eqnarray}
U^{(i3)}&=&{\cal C}_F[I(p_1,p_2) - I(p_1,k_2)]
\nonumber \\
& &-\frac{1}{2 N_c}[I(p_1,p_3) +I(k_1,p_3)+
I(k_2,p_3) -I(p_2,p_3)]\;,
\label{ui3}
\end{eqnarray}
according to Fig.~3(b). We present only the explicit expression of 
$I(p_1,p_3)$,
\begin{eqnarray}
I(p_1,p_3) & = & -\frac{\alpha_{s}}{\pi}
          \frac{1}{\epsilon} \left[
\frac{1}{2} \ln\frac{(p_1 \cdot p_3)^2}{p_1^2}
-\frac{1}{2}\ln\frac{(p_3 \cdot n)^2}{n^2}
-f(p_1,n) +1 \right]\;.
\label{uhl}
\end{eqnarray}
All other $I$'s can be derived simply from Eqs.~(\ref{uhh}), (\ref{ull})
and (\ref{uhl}) with appropriate replacement of the kinematic variables.
Following the similar procedures, we have
\begin{eqnarray}
\gamma^{(i1)}_U&=&-\frac{\alpha_{s}}{\pi}\frac{1}{N_c}\;,
\nonumber \\
\gamma^{(i2)}_U&=&-\frac{\alpha_{s}}{\pi}\left[
\frac{1}{N_c}\left(\frac{2r^2}{r^4-1}\ln r+1\right)-
{\cal C}_F\frac{2r^2}{r^2+1}\ln r \right]\;,
\nonumber \\
\gamma^{(i3)}_U&=&-\frac{\alpha_{s}}{\pi}\left[
\frac{1}{N_c}\left(\frac{r^2}{r^2+1}\ln r+1\right)-
{\cal C}_F\frac{4r^2}{r^4-1}\ln r \right]\;.
\end{eqnarray}
\vskip 1.0cm

\centerline{\bf III. FACTORIZATION FORMULAS}
\vskip 0.5cm

The decay rate of $B^-\to D^0\pi^-$ is \cite{CL}
\begin{equation}
\Gamma=\frac{1}{128\pi}G_F^2|V_{cb}|^2|V_{ud}|^2M_B^3\frac{(1-r^2)^3}{r}
|{\cal M}|^2\;,
\label{dr}
\end{equation}
with the decay amplitude
\begin{eqnarray}
{\cal M}=f_\pi[(1+r)\xi_+-(1-r)\xi_-]+
f_D\xi_i+{\cal M}_e+{\cal M}_i\;,
\label{M1}
\end{eqnarray}
$f_{D}$ and $f_\pi$ being the $D$ meson and pion decay constant,
respectively. 
The form 
factors $\xi_\pm$ associated with the external $W$ emissions from 
Figs.~2(a)-2(d) are given by 
\begin{eqnarray}
\xi_+&=& 16\pi{\cal C}_F\sqrt{r}M_B^2
\int_{0}^{1}d x_{1}d x_{2}\,\int_{0}^{\infty} b_1d b_1 b_2d b_2\,
\phi_B(x_1)\phi_{D}(x_2)\alpha_s(t_e)
a_1(t_e)
\nonumber \\
& &\times 
[(1+\zeta_+x_2r)h_e(x_1,x_2,b_1,b_2,m_e)
+(r+\zeta_+x_1)h_e(x_2,x_1,b_2,b_1,m_e)]
\nonumber \\
& &\times \exp[-S_B(t_e)-S_{D}(t_e)-S_U^{(e)}(t_e)]\;,
\label{+}\\
\xi_-&=& 16\pi{\cal C}_F\sqrt{r}M_B^2
\int_{0}^{1}d x_{1}d x_{2}\,\int_{0}^{\infty} b_1d b_1 b_2d b_2\,
\phi_B(x_1)\phi_{D}(x_2)\alpha_s(t_e)a_1(t_e)
\nonumber \\
& &\times \zeta_- [x_2rh_e(x_1,x_2,b_1,b_2,m_e)
-x_1 h_e(x_2,x_1,b_2,b_1,m_e)]
\nonumber \\
& &\times \exp[-S_B(t_e)-S_{D}(t_e)-S_U^{(e)}(t_e)]\;,
\label{-}
\end{eqnarray}
with the constants \cite{L1}
\begin{eqnarray}
\zeta_+&=&\frac{1}{2}\left[\eta-\frac{3}{2}+
\sqrt{\frac{\eta-1}{\eta+1}}\left(\eta-\frac{1}{2}\right)\right]\;,
\nonumber \\
\zeta_-&=&-\frac{1}{2}\left[\eta-\frac{1}{2}+
\sqrt{\frac{\eta+1}{\eta-1}}\left(\eta-\frac{3}{2}\right)\right]\;.
\end{eqnarray}
$\eta=(1+r^2)/(2r)$ is the maximal velocity transfer involved in the
decay process. The form factors $\xi_i=\xi_{i1}+\xi_{i23}$ associated with 
the internal $W$ emissions are 
\begin{eqnarray}
\xi_{i1}&=&16\pi{\cal C}_F\sqrt{r}M_B^2
\int_0^1 dx_1dx_3\int_0^{\infty}b_1db_1b_3db_3
\phi_B(x_1)\phi_\pi(x_3)\alpha_s(t_i)c_2(t_i)
\nonumber \\
& &\times \left[[1+x_3(1-r^2)]h_i(x_1,x_3,b_1,b_3,m_i)
+x_1r^2 h_i(x_3,x_1,b_3,b_1,m_i)\right]
\nonumber \\
& &\times\exp[-S_B(t_i)-S_\pi(t_i)-S_U^{(i1)}(t_i)]\;,
\label{int1} 
\end{eqnarray}
from Figs.~2(a) and 2(b), and
\begin{eqnarray}
\xi_{i23}&=&16\pi{\cal C}_F\sqrt{r}M_B^2
\int_0^1 dx_1dx_3\int_0^{\infty}b_1db_1b_3db_3
\phi_B(x_1)\phi_\pi(x_3)\alpha_s(t_i)\frac{c_1(t_i)}{N}
\nonumber \\
& &\times \left\{[1+x_3(1-r^2)]h_i(x_1,x_3,b_1,b_3,m_i)
\exp[-S_U^{(i2)}(t_i)]\right.
\nonumber \\
& &\left.+x_1r^2 h_i(x_3,x_1,b_3,b_1,m_i)\exp[-S_U^{(i3)}(t_i)]
\right\}
\nonumber \\
& &\times\exp[-S_B(t_i)-S_\pi(t_i)]\;,
\label{int} 
\end{eqnarray}
from Figs.~2(c) and 2(d). Strickly speaking, $\xi_i$ should not be 
classified as a factorizable contribution because of the irreducible
soft gluons that attach the $B$ and $D$ mesons and the pion and $D$ meson.
Only at the lowest order of soft corrections, {\it i.e.} $S_U=0$, is it a 
factorizable contribution. We shall consider this point, when
identifying the nonfactorizable contributions below.

The exponents 
\begin{eqnarray}
S_B(\mu)&=&s(x_1P_1^-,b_1)+2\int_{1/b_1}^{\mu}\frac{d{\bar \mu}}{\bar \mu}
\gamma(\alpha_s({\bar \mu}))\;,
\nonumber \\
S_{D}(\mu)&=&s(x_2P_2^+,b_2)+s((1-x_2)P_2^+,b_2)+2\int_{1/b_2}^{\mu}
\frac{d{\bar \mu}}{\bar \mu}\gamma(\alpha_s({\bar \mu}))\;,
\nonumber \\
S_\pi(\mu)&=&s(x_3P_3^-,b_3)+s((1-x_3)P_3^-,b_3)+
2\int_{1/b_3}^{\mu}\frac{d{\bar \mu}}{\bar \mu}
\gamma(\alpha_s({\bar \mu}))\;,
\label{wpe}
\end{eqnarray}
correspond to the summation of the reducible radiative corrections grouped 
into the $B$ meson wave function $\phi_B$, the $D$ meson wave function 
$\phi_D$, and the pion wave function $\phi_\pi$, respectively. The quark 
anomalous dimension $\gamma=-\alpha_s/\pi$, is related to
$\gamma_\phi=2\gamma$ introduced before. The exponents
\begin{eqnarray}
S_U^{(j)}(\mu)=\int_{w}^{\mu} {d\bar{\mu} \over \bar{\mu}}  
\gamma_{U}^{(j)}(\alpha_s({\bar \mu}))\;,
\label{su}
\end{eqnarray}
for $j=e,i1,i2$, and $i3$, correspond to the summation of the irreducible 
corrections grouped into the soft function $U$, where the lower bound 
$w=\min(1/b_i)$ is chosen to collect the largest soft logarithms. The wave 
functions satisfy the normalization 
\begin{equation}
\int_0^1\phi_{B,D,\pi}(x)dx=\frac{f_{B,D,\pi}}{2\sqrt{6}}\;,
\end{equation}
$f_B$ being the $B$ meson decay constant.

In Eqs.~(\ref{+}), (\ref{-}), (\ref{int1}) and (\ref{int}) the functions 
$h$'s, the Fourier transform of the lowest-order $H$ from Figs.~2(a)-2(d),
are given by
\begin{eqnarray}
h_e(x_1,x_2,b_1,b_2,m_e)&=&K_{0}\left(\sqrt{x_1x_2m_e}b_1\right)
\nonumber \\
& &\times \left[\theta(b_1-b_2)K_0\left(\sqrt{x_2m_e}
b_1\right)I_0\left(\sqrt{x_2m_e}b_2\right)\right.
\nonumber \\
& &\left.+\theta(b_2-b_1)K_0\left(\sqrt{x_2m_e}b_2\right)
I_0\left(\sqrt{x_2m_e}b_1\right)\right],
\label{dh}\\
h_i(x_1,x_3,b_1,b_3,m_i)&=&h_e(x_1,x_3,b_1,b_3,m_i)\;,
\label{hint}
\end{eqnarray}
with $m_e=M_B^2$ and $m_i=(1-r^2)M_B^2$. The hard scales $t$'s take the 
maximum of all energies involved in $H$:
\begin{eqnarray}
t_e&=&{\rm max}(\sqrt{x_1m_e},\sqrt{x_2m_e},1/b_1,1/b_2)
\nonumber \\
t_i&=&{\rm max}(\sqrt{x_1m_i},\sqrt{x_3m_i},1/b_1,1/b_3)\;. 
\label{tei}
\end{eqnarray}
It is an important feature that the Sudakov form factor $e^{-s}$ exhibits a 
strong suppression in the large $b$ region. Hence, Sudakov suppression 
guarantees that the main contributions arise from the large $t$ region, where 
the running coupling constant $\alpha_s(t)$ is small, and perturbation theory 
is relatively reliable.

The factorization formulas for the nonfactorizable external and internal 
$W$-emission amplitudes ${\cal M}_e$ and ${\cal M}_i$, respectively, 
contain the kinematic variables of all the three mesons. The integration 
over $b_3$ can be performed trivially, leading to $b_3=b_1$ or $b_3=b_2$. 
Their expressions are
\begin{eqnarray}
{\cal M}_e&=& 32\pi\sqrt{2N}{\cal C}_F\sqrt{r}M_B^2 
\int_0^1 [dx]\int_0^{\infty}
b_1 db_1 b_2 db_2
\phi_B(x_1)\phi_{D}(x_2)\phi_\pi(x_3) \nonumber \\
& &\times \biggl\{ \alpha_s(t_e^{(1)})\frac{c_2(t_e^{(1)})}{N}  
\exp[-S(t_e^{(1)})|_{b_3=b_2}-S_U^{(e)}(t_e^{(1)})]
\nonumber \\
& &\hspace{0.5cm}\times
[(1-r^2)(1-x_3)-x_1+(r-r^2)(x_1-x_2)]h^{(1)}_e(x_i,b_i) 
\nonumber \\
& &\hspace{0.5cm}
- \alpha_s(t_e^{(2)})\frac{c_2(t_e^{(2)})}{N}  
\exp[-S(t_e^{(2)})|_{b_3=b_2}-S_U^{(e)}(t_e^{(2)})]
\nonumber \\
& &\hspace{0.5cm}
\times[(2-r)x_1-(1-r)x_2-(1-r^2)x_3]h^{(2)}_e(x_i,b_i) \biggr\}\;,
\label{mb}\\
{\cal M}_i&=& 32\pi\sqrt{2N}{\cal C}_F\sqrt{r}M_B^2
\int_0^1 [dx]\int_0^{\infty}b_1 db_1 b_2 db_2
\phi_B(x_1)\phi_{D}(x_2)\phi_\pi(x_3) \nonumber \\
& &\times 
\biggl\{ \alpha_s(t_i^{(1)})\frac{c_1(t_i^{(1)})}{N}  
\exp[-S(t_i^{(1)})|_{b_3=b_1}-S_U^{(i2)}(t_i^{(1)})]
\nonumber \\
& &\hspace{0.5cm}
\times[x_1-x_2-x_3(1-r^2)]h^{(1)}_i(x_i,b_i) 
\nonumber \\
& &\hspace{0.5cm}
+ \alpha_s(t_i^{(2)})\frac{c_1(t_i^{(2)})}{N}  
\exp[-S(t_i^{(2)})|_{b_3=b_1}-S_U^{(i3)}(t_i^{(2)})]
\nonumber \\
& &\hspace{0.5cm}
\times[(x_1+x_2)(1+ r^2)-1]h^{(2)}_i(x_i,b_i) 
\biggr\}\;,
\label{md}
\end{eqnarray}
with $[dx]\equiv dx_1dx_2dx_3$ and $S=S_B+S_{D}+S_\pi$. The functions
$h^{(j)}$, $j=1$ and 2, appearing in Eqs.~(\ref{mb}) and (\ref{md}), are
derived from Figs.~2(e) and 2(f):
\begin{eqnarray}
\everymath{\displaystyle}
h^{(j)}_e&=& \left[\theta(b_1-b_2)K_0\left(BM_B
b_1\right)I_0\left(BM_Bb_2\right)\right. \nonumber \\
& &\quad \left.
+\theta(b_2-b_1)K_0\left(BM_B b_2\right)
I_0\left(BM_B b_1\right)\right]\;  \nonumber \\
&  & \times \left( \begin{array}{cc}
 K_{0}(B_{j}M_Bb_{2}) &  \mbox{for $B_{j} \geq 0$}  \\
 \frac{i\pi}{2} H_{0}^{(1)}(|B_{j}|M_Bb_{2})  & \mbox{for $B_{j} \leq 0$}
  \end{array} \right)\;,           
\\
\everymath{\displaystyle}
h^{(j)}_i&=& \left[\theta(b_1-b_2)K_0\left(DM_B
b_1\right)I_0\left(DM_Bb_2\right)\right. \nonumber \\
& &\quad \left.
+\theta(b_2-b_1)K_0\left(DM_B b_2\right)
I_0\left(DM_B b_1\right)\right]\;  \nonumber \\
&  & \times \left( \begin{array}{cc}
 K_{0}(D_{j}M_Bb_{2}) &  \mbox{for $D_{j} \geq 0$}  \\
 \frac{i\pi}{2} H_{0}^{(1)}(|D_{j}|M_Bb_{2})  & \mbox{for $D_{j} \leq 0$}
  \end{array} \right)\;,           
\label{hjd}
\end{eqnarray}
with the variables
\begin{eqnarray}
B^{2}&=&x_{1}x_{2}(1-r^{2})\;,
\nonumber \\
B_{1}^{2}&=&(x_{1}-x_{2})x_{3}(1-r^{2})+x_{1}x_{2}(1+r^{2})\;,
\nonumber \\
B_{2}^{2}&=&x_{1}x_{2}(1+r^{2})-(x_{1}-x_{2})(1-x_{3})(1-r^{2})\;,
\nonumber \\
D^{2}&=&x_{1}x_{3}(1-r^{2})\;, 
\nonumber \\
D_{1}^{2}&=&(x_{1}-x_{2})x_{3}(1-r^{2})\;,
\nonumber \\
D_{2}^{2}&=&(x_{1}+x_{2})r^{2}-(1-x_{1}-x_{2})x_{3}(1-r^{2})\;.
\end{eqnarray}
Similarly, the scales $t^{(j)}$ are chosen as 
\begin{eqnarray}
t_e^{(j)}&=&{\rm max}(BM_B,|B_j|M_B,1/b_1,1/b_2)\;,
\nonumber \\
t_i^{(j)}&=&{\rm max}(DM_B,|D_j|M_B,1/b_1,1/b_2)\;.
\label{tjj}
\end{eqnarray}
\vskip 2.0cm

\centerline{\bf IV. RESULTS AND DISCUSSIONS}
\vskip 0.5cm

In the evaluation of the various form factors and amplitudes, we adopt
$G_F=1.16639\times 10^{-5}$ GeV$^{-2}$, the decay constants $f_B=200$ MeV, 
$f_D=220$ MeV, and $f_\pi=132$ MeV \cite{A}, the CKM matrix elements 
$|V_{cb}|=0.040$ and $|V_{ud}|=0.974$, the masses $M_B=5.28$ GeV and 
$M_D=1.87$ GeV \cite{PDG}, the $B^-$ meson lifetime $\tau_{B^-}=1.68$ ps 
\cite{B}, the $B$ and $D$ meson wave functions \cite{ASY}, 
\begin{equation}
\phi_{B,D}(x)=\frac{N_{B,D}}{16\pi^2}\frac{x(1-x)^2}
{M_{B,D^{(*)}}^2+C_{B,D}(1-x)}\;,
\label{bw}
\end{equation}
and the Chernyak-Zhitnitsky pion wave function \cite{CZ},
\begin{eqnarray}
\phi_\pi(x)=\frac{5\sqrt{6}}{2}f_\pi x(1-x)(1-2x)^2\;.
\label{pwf}
\end{eqnarray}
The normalization constants $N_B=590.8$ GeV$^3$ and $N_D=92.85$ GeV$^3$,
and the shape parameters $C_B=-27.6$ GeV$^2$ and $C_D=-3.372$ GeV$^2$
are determined by fitting the predictions from Eq.~(\ref{dr}) to the data
of the decay rates of $B\to D^{(*)}\pi$ \cite{A}.

Our formalism can be applied to the decay $D^+\to {\bar K}^{0}\pi^+$ 
directly, which occurs through a similar effective Hamiltonian
but with the four-fermion operators
$O_{1}  =  (\bar{s}_{L}\gamma_{\mu} c_{L})
 (\bar{d}_{L}\gamma^{\mu} u_{L})$ and
$O_{2}  =  (\bar{d}_{L}\gamma_{\mu} c_{L})            
(\bar{s}_{L}\gamma^{\mu} u_{L})$. The expression of the decay rate $\Gamma$
is also similar to Eq.~(\ref{dr}), but with the CKM matrix element 
$|V_{cs}|=1.0$ substituted for $|V_{cb}|$, and the meson masses $M_D$ and 
$M_K=0.497$ GeV \cite{PDG2} for $M_B$ and $M_{D}$, respectively. In all the 
form factors and amplitudes the kinematic variables of the 
$B$ $(D)$ meson are replaced by those of the $D$ $(K)$ meson. The $D^+$ 
meson lifetime is $\tau_{D^+}=1.05$ ps \cite{PDG2}, and the $D$ meson wave 
function has been defined above. For the kaon, we adopt the decay constant 
$f_K=160$ MeV, and the wave function \cite{CZ},
\begin{eqnarray}
\phi_K(x)&=&\frac{\sqrt{6}}{2}f_Kx(1-x)[3.0(1-2x)^2+0.4]\;.
\label{kw0}
\end{eqnarray}

Because of the smaller $D$ meson mass, the transverse degrees of freedom
are more important in the definitions of the hard scales $t$. Hence, we
choose the maximum of the scales $1/b_i$ for the arguments $t$ of the Wilson 
coefficients. In this case Sudakov suppression is weaker, and 
insufficient to diminish the contributions from the region with $t$ close 
to $\Lambda_{\rm QCD}$, where $c_{1,2}(t)$ diverge. To have meaningful 
predictions, we choose $t=\max(1/b_i,t_c)$ in the numerical analysis,
$t_c=(1+\epsilon)\Lambda_{\rm QCD}$, such that the Wilson coefficients
are frozen at $c_{1,2}(t_c)$ as $\max(1/b_i) \le t_c$.
The parameter $\epsilon=0.0000227$ is determined
from the data of the decays $D\to K^{(*)}\pi$ \cite{PDG2}. 

After including the soft gluons, the factorizable external $W$-emission
contributon $\chi^{(f)}_e$ and the factorizable internal $W$-emission
contributon $\chi^{(f)}_i$ to the decay amplitude ${\cal M}$ in 
Eq.~(\ref{M1}) should be identified as
\begin{eqnarray}
\chi^{(f)}_e&=&f_\pi[(1+r)\xi_+-(1-r)\xi_-]\;,
\nonumber \\
\chi^{(f)}_i&=&f_{D(K)}\xi_i|_{S_U=0}\;,
\end{eqnarray}
where $\xi_{i}|_{S_U=0}$ denotes the lowest order internal $W$-emission 
contributions obtained from Figs.~2(a)-2(d) by turning off the soft
corrections. The nonfactorizable external $W$-emission contributon
$\chi^{(nf)}_e$ and the nonfactorizable internal $W$-emission contributon
$\chi^{(nf)}_i$  are then
\begin{eqnarray}
\chi^{(nf)}_e&=&{\cal M}_e\;,
\nonumber \\
\chi^{(nf)}_i&=&f_{D(K)}(\xi_{i}-\xi_{i}|_{S_U=0})+{\cal M}_i\;.
\label{nf}
\end{eqnarray}
We shall show that our predictions of $\chi_{1,2}$ are positive and negative
for the $B$ and $D$ meson decays, respectively, consistent with the
phenomenological arguments in Eq.~(\ref{chii}) \cite{HYC1,CT}. However, the
positive nonfactorizable contribution $\chi_1$ for bottom decays is
contrary to the QCD sum rule predictions \cite{BS93,IH,KR},
which we shall comment on later.

For $r=M_D/M_B=0.354$, we have the anomalous dimensions 
$\gamma_U^{(e)}=-0.421\alpha_s/\pi$,
$\gamma_U^{(i1)}=-0.333\alpha_s/\pi$, $\gamma_U^{(i2)}=-0.730\alpha_s/\pi$,
and $\gamma_U^{(i3)}=0.413\alpha_s/\pi$. For $r=M_K/M_D=0.265$, we have 
$\gamma_U^{(e)}=-0.396\alpha_s/\pi$, $\gamma_U^{(i1)}=-0.333\alpha_s/\pi$,
$\gamma_U^{(i2)}=-0.628\alpha_s/\pi$, and 
$\gamma_U^{(i3)}=0.196\alpha_s/\pi$. The results of the various form factors
and amplitudes for the decays $B^-\to D^{0}\pi^-$ and 
$D^+\to {\bar K}^{0}\pi^+$ are exhibited in Table I. The rows entitled by
$S_U\not =0$, whose values match the data, are derived with the soft
corrections taken into account. Those by $S_U=0$ only help to extract the
nonfactorizable contributions $\chi^{(nf)}$, and to investigate the
importance of the soft corrections. Because the $D$ meson is lighter, the
scales $t$ can run to a lower value as indicated by Eqs.~(\ref{tei}) and
(\ref{tjj}). The exponentials $e^{-S_U}$, which basically act as 
enhancing factors, then amplify the contributions from this region with
smaller $t$, where the Wilson coefficients are larger. Therefore, the soft
gluon effects are more important in charm decays as shown in Table I.

The factorizable external $W$-emission contributions are positive in both
the $B$ and $D$ meson decays, and their magnitudes increase, after including
the soft corrections with $\gamma_U^{(e)}<0$. $\xi_{i}$ changes sign, since
the Wilson coefficient $c_2$ in $\xi_{i1}$ becomes so negative, when evolving
from the characteristic scale of the $B$ meson decay to that of the $D$
meson decay, that it overcomes the positive $c_1/N_c$ in $\xi_{i23}$. Their
magnitudes also increase because of $\gamma_U^{(i1)},\; \gamma_U^{(i2)}<0$.
The real parts of the nonfactorizable amplitudes ${\cal M}_e$ (${\cal M}_i$)
are always negative (positive) due to the negative $c_2$ (positive
$c_1/N_c$) in the $B$ and $D$ meson decays. At first sight, these
observations differ from the phenomenological extractions
in Eq.~(\ref{chii}) \cite{HYC1,CT}. As argued before, the nonfactorizable
contributions should be appropriately identified according to
Eq.~(\ref{nf}), whose results are listed in Table I.
We have $Re(\chi^{(nf)}_i)=+0.0189$ GeV for the $B$ meson decay and
$Re(\chi^{(nf)}_i)=-0.4715$ GeV for the $D$ meson decay.
It is easy to attribute the sign change of $\chi^{(nf)}_i$ 
to the stronger enhancement of $\xi_i$ by the soft corrections in
charm decays.

To compare our predictions with Eq.~({\ref{chii}), we present in Table I
the BSW parameters $a_1$, $a_2$, $c_2\chi_1$, and $c_1\chi_2$,
corresponding to $\chi^{(f)}_e$, $\chi^{(f)}_i$, $\chi^{(nf)}_e$, and
$\chi^{(nf)}_i$, respectively. The signs, except for those of
$Re(\chi^{(nf)}_e)$ and $c_2\chi_1$ associated with the $D$ meson decay, are
consistent. The discrepancy between the signs of $Re(\chi^{(nf)}_e)$ and
$c_2\chi_1$ for the $D$ meson decay will be resolved, if two-loop soft
functions $U$ are considered. At this order, two soft gluons attach the
valence quark lines of, for example, the $B$ meson and the pion in Figs.~2(a)
and 2(b), for which the associated color traces do not vanish. Then the form
factors $\xi_\pm$ should be classified as being nonfactorizable, and
$\chi^{(nf)}_e$ be redefined by a similar expression to Eq.~(\ref{nf}). It
is expected based on Table I that $Re(\chi^{(nf)}_e)$ for the $B$ meson
decay remains negative, while that for the $D$ meson decay changes sign.

At last, we come to the comparision of our predictions with those derived
from QCD sum rules \cite{BS93}. The conclusion that the nonfactorizable
contribution $\chi_1$ in bottom decays is negative holds only in the
kinematic limit $M_B$, $M_D\to \infty$ but with $\Delta M=M_B-M_D$ fixed
\cite{BS93}. In this limit both the $B$ and $D$ mesons are at rest, and soft
corrections should be more important. The reason is similar to that for the
larger soft effects in charm decays given before. For extremely heavy mesons,
the wave functions peak at $x\to 0$ \cite{L1}, and the characteristic scales
$t$ in Eq.~(\ref{tei}) becomes smaller. Then the form factors $\xi$ may
increase by a significant factor, since the contributions from the region
with large Wilson coefficients are enhanced by the exponentials $e^{-S_U}$.
Following Eq.~(\ref{nf}), it is possible that the sign of $\chi^{(nf)}$ for
bottom decays is reversed. However, we argue that the above kinematic
condition is inappropriate for realistic $B$ meson decays, and thus the
conclusion in \cite{BS93} is in doubt.

In this paper we have not only completed the Sudakov resummation up to
next-to-leading logarithms, but also investigated the soft gluon effects
in heavy meson decays based on the three-scale PQCD factorization
theorem. By identifying the nonfactorizable contributions carefully, we are
able to explain the variation and especially the sign changes of the 
amplitudes in bottom and charm decays. Certainly, these subjects still
need more thorough studies in order to have a full understanding of their
dynamical origin.

\vskip 0.5cm
We thank H.Y. Cheng for a helpful discussion, and T.W. Yeh for his
assistance in the numerical analysis.
This work was supported by the National Science Council of Republic
of China under the Grant No. NSC-86-2112-M-194-007 and 
NSC-86-2112-M-002-010-Y.

\newpage

TABLE I. The various form factors and amplitudes for the decays
$B^-\to D^{0}\pi^-$ and $D^+\to {\bar K}^{0}\pi^+$. 
The unit is $10^{-3}$ GeV.

\vskip 1.0cm
\[\begin{array}{ccccc}\hline\hline
B\to D\pi&{\rm external}\;W&f_D\xi_{i}&{\cal M}_e&{\cal M}_i\\ 
         &{\rm (factorizable)}&       &          &          \\ 
\hline
S_U=0    &106.5 &2.5 &-5.8+19.8i &18.5-12.5i\\
S_U\not=0 &108.5 &2.6 &-5.8+20.0i &18.8-11.0i\\
\hline \hline
D\to {\bar K}\pi&{\rm external}\;W&f_K\xi_{i}&{\cal M}_e&{\cal M}_i\\ 
         &{\rm (factorizable)}&       &          &          \\ 
\hline
S_U=0    &267.0 &-21.3 &-18.8+19.0i &37.8-36.7i\\
S_U\not=0 &1075.0 &-529.3 &-18.5+13.6i &36.5-30.3i\\
\hline\hline
         &\chi^{(f)}_e &\chi^{(f)}_i &\chi^{(nf)}_e &\chi^{(nf)}_i \\
         & (a_1)       & (a_2)       & (c_2\chi_1)    & (c_1\chi_2)  \\
\hline
B\to D\pi & 108.5 & 2.5 & -5.8+20.0i & 18.9-11.0i \\
          & (1.03) & (0.11) & (-0.01) & (0.12) \\
D\to {\bar K}\pi & 1075.0 & -21.3 & -18.5+13.6i & -471.5-30.3i \\
                 & (1.09) & (-0.1) & (0.19) & (-0.45) \\
\hline\hline

\end{array}\]

\newpage
\centerline{\large \bf Figure Captions}
\vskip 0.5cm

\noindent
{\bf Fig. 1.} (a) $O(\alpha_s)$ factorization of infrared and hard 
contributions. (b) $O(\alpha_s)$ factorization into a ``harder" function,
a soft function and a hard decay subamplitude. 
\vskip 0.5cm

\noindent 
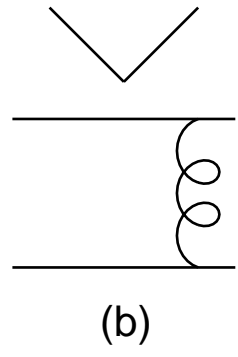
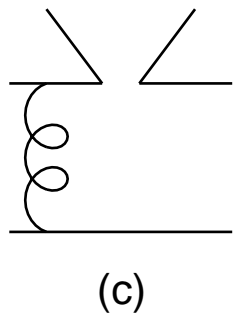
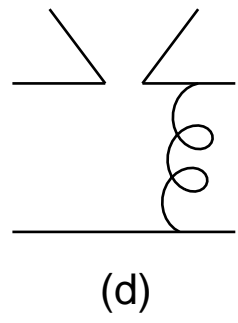
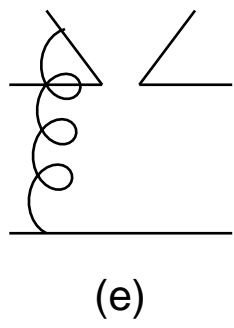
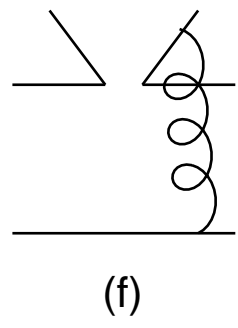
{\bf Fig. 2.} Lowest-order diagrams for the hard subamplitude.
\vskip 0.5cm

\noindent 
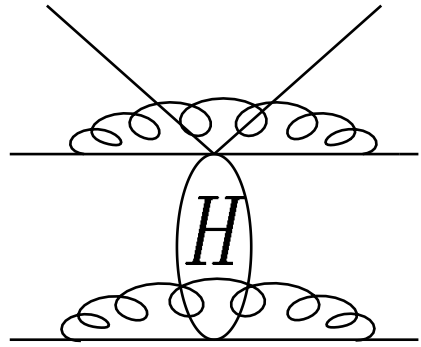
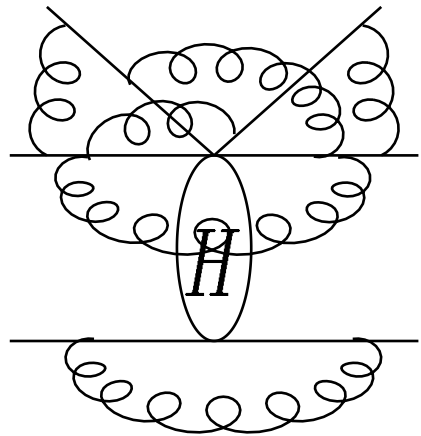
{\bf Fig. 3.} Soft gluon corrections to Fig.~2.

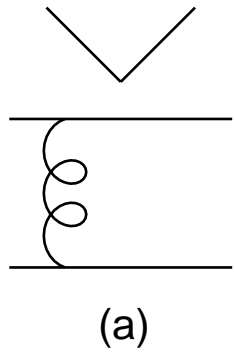
\end{document}